\documentclass[prx,aps,amssymb,twocolumn,reprint]{revtex4-1}

\usepackage{amsmath}    

\usepackage{amsfonts}
\usepackage{amssymb}
\usepackage[colorlinks=true,citecolor=blue,linkcolor=red]{hyperref}
\usepackage{graphicx}
\usepackage[dvipsnames]{xcolor}

\usepackage{ulem}
\usepackage{bm} 
\usepackage{subfigure}
\usepackage{dcolumn}
\usepackage{mathtools}

\usepackage{array,booktabs,ragged2e}	
\usepackage{soul}						% strike out text using \st{}
\usepackage{cancel}						% detto, but more clearly visible when reading
\usepackage{tabularx}% for table in the CHS-model discussion
\usepackage{multirow}
\usepackage{hhline}
\usepackage{adjustbox}
\usepackage{sidecap}
\usepackage{threeparttable}

\usepackage[utf8]{inputenc}
\usepackage[T1]{fontenc}

\newcommand{\MR}{{\rm MR}}

\newcolumntype{P}[1]{>{\raggedleft\arraybackslash}p{#1}}
\newcolumntype{R}[1]{>{\centering\arraybackslash}p{#1}}

\begin{document}

%%%%%%%%%%%%%%%%%%%%%%%%%
%%% TITLE INFORMATION %%%
%%%%%%%%%%%%%%%%%%%%%%%%%
\title{The inadequacy of the $\rho$-T curve for phase transitions in the presence of magnetic fields}

\author{Shengnan Zhang$^{1}$}
\author{Zhong Fang$^{1,2,3}$}
\author{Hongming Weng$^{1,2,3}$}
%\email{hmweng@iphy.ac.cn}
\author{Quansheng Wu$^{1,2}$}
\email{quansheng.wu@iphy.ac.cn}

\affiliation{${^{1}}$Beijing National Laboratory for Condensed Matter Physics and Institute of physics,
Chinese academy of sciences, Beijing 100190, China}
\affiliation{${^{2}}$University of Chinese academy of sciences, Beijing 100049, China}
\affiliation{${^{3}}$Songshan Lake Materials Laboratory, Dongguan, Guangdong 523808, China}
\date{\today}

\begin{abstract}
The $\rho(T)$ curve is traditionally employed to discern metallic, semiconductor, and insulating behaviors in materials, with any deviations often interpreted as indicative of phase transitions. However, does this interpretation hold under the influence of a magnetic field? Our research addresses this critical question by reevaluating the $\rho(T)$ curve in the presence of magnetic field. We uncover that metal-insulator shifts and reentrant metallic states may not indicate true phase transitions but rather originate from the scaling behavior of magnetoresistance, influenced by magnetic field and temperature through a power-law dependence. Employing advanced first-principles calculations and the Boltzmann method, we analyzed the magnetoresistance of SiP$_2$ and NbP across a range of conditions, successfully explaining not only the reentrant behavior observed in experiments but also resolving the discrepancies in magnetoresistance behavior reported by different research groups. These findings challenge the conventional use of the $\rho(T)$ curve as a straightforward indicator of phase transitions under magnetic conditions, highlighting the essential need to exclude typical magnetoresistance effects due to the Lorentz force before confirming such transitions. This novel insight reshapes our understanding of complex material properties in magnetic fields and sets a new precedent for the interpretation of transport phenomena in condensed matter physics.
\end{abstract} 

\maketitle

Typically resistivity-temperature curve $\rho(T)$ serve as an empirical criterion to determine the phase of a material according to solid physics\cite{Mermin1976,ziman}. Metals exhibit an increase in resistivity with rising temperature, while insulators show a decrease, and semiconductors demonstrate a decrease at low temperatures but increase at higher temperatures. At the beginning of this century, theoretical and experimental investigations into graphite and bismuth spurred detailed studies into their temperature dependent magnetotransport propeties\cite{graphiteMR1, graphiteMR2, graphiteMR3, reen_grap1, graphiteMR4, graphiteMR5, grap_scaling, reen_grap2, reen_grap3,  reen_grap4,  graphiteMR6}. Notably, Y. Kopelevich et al. observed a reentrant metallic behavior in graphite, potentially associated with quantum Hall effect and superconducting correlations induced by Landau-level quantization \cite{reen_grap1}. In 2005, Du et al. demonstrate that the magnetic-field-induced metal-insulator transition of graphite and bismuth is facilitated by a unique ordering and spacing of three characteristic energy scales\cite{grap_scaling}. 

Since the discovery of extreme magnetoresistance (XMR) in WTe$_2$ in 2014~\cite{MRWTe2nature}, interest in magnetoresistance(MR) and magnetic field-induced metal-insulator transitions has been rekindled. A study in 2015 revisiting Kohler’s rule shed light on why resistivity, under the influence of a magnetic field, mirrors characteristics similar to metal-insulator transitions as temperature varies~\cite{WTe2Kohler}. This phenomenon, known as the upturn phenomenon, has subsequently been observed in numerous materials exhibiting XMR~\cite{turnonCd3As2,turnonLaSb,turnonNbP, turnonMoTe2,turnonLaBi,turnonaWP2,turnonbWP2,turnonMoO2, turnonSiP2, turnonReO3,turnonTaSe3,turnonInBi,turnonGdBi, turnonWTe2Ni,turnonTaP,turnonCaCdSn}. In 2017, reentrant metallic behavior was observed in the Weyl semimetal NbP, as reported in Ref.\cite{reentrantNbP}, where a semiclassical theory was utilized to explain these anomalies without resorting to exotic mechanisms.

However the behavior of MR dependent on magnetic field and temperature is extremely complex, currently lacking a theoretical framework to identify the dominant mechanisms and link them to experimental observations. In this study, we systematically investigate the scaling behavior of resistivity under various magnetic fields and temperatures. Utilizing the two-band model and examining representative materials NbP and SiP$_2$,we analyze the different MR behaviors and elucidate the phenomena of "reentrant metallic" and "metal-insulator-like". In order to calculate the magnetotransport properties for real materials from the combined first-principles calculation and semiclassical Boltzmann transport theory, we construct tight-binding models using the first principles calculation software such as VASP~\cite{Vasp1, Vasp2} along with Wannier function techniques such as Wannier90~\cite{NMwannier90}. Our detailed computational methods are thoroughly explained in Ref.~\cite{MRZhangprb, Liuyiprb} and implemented within the WannierTools software package~\cite{WUWT}.

\begin{figure}[htpb]
    \centering
    \includegraphics[width=1\linewidth]{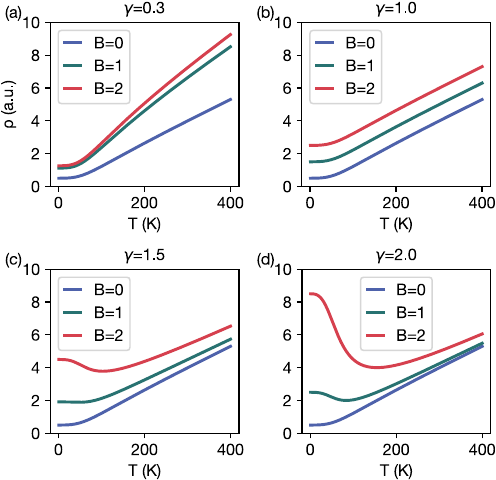}
    \caption{Temperature dependent resistivity at different magnetic field for different exponent $\gamma$.}
    \label{fig1}
\end{figure}

According to Kohler's rule~\cite{Kohlerrule} and Chambers equation\cite{Mermin1976,Chambers1952}, the MR can be described as a function of the product of magnetic field $B$ and relaxation time $\tau$\cite{supp}, 
\begin{align}
    {\rm MR} = \frac{\rho(B)-\rho_0}{\rho_0}
    \propto (B\tau)^{\gamma}= (\frac{B}{\alpha\rho_0})^{\gamma}
    \label{equ1}
\end{align}
where $\tau$ is approximated as $\tau=1/\alpha\rho_0$ with $\rho_0$ is the temperature dependent resistivity at $B=0$. Then the longitudinal resistivity $\rho(B,T)$ can be expressed as,
\begin{align}
        \rho(B,T)=\rho_0(T)+\zeta \frac{B^{\gamma}}{(\alpha\rho_0)^{\gamma-1}}
        \label{equ2}
\end{align}
where $\zeta$ is a material dependent constant.  Therefore the exponent $\gamma$ in the power-law dependence of MR on the magnetic field is particularly critical, since it determines the potential competition between the first and second term on the right side of Eq.(\ref{equ2}), leading to a revert in the temperature dependence of the resistivity beyond a certain magnetic field strength. We adopted the Kohler's rule approach to extract the main physics picture, but the conclusions are robust, which is explained in the end of this article. 

To begin with, we consider the simplest case, i.e.$\gamma = 1$, where the second term on the left side of Eq.(\ref{equ2}) stabilizes to a constant, rendering the expression for $\rho(B,T)$ as $\rho = \rho_0(T) + \text{const}$. Consequently, $\rho$ precisely inherits the temperature dependence of $\rho_0$, adjusted by a constant shift. Thus, one can easily visualize a series of $\rho(T)$ curves that are evenly spaced across different temperatures and varying magnetic fields, as illustrated in Fig. \ref{fig1}(b). Note that the zero-field resistivity $\rho_0$ is calculated by the  Bloch-Grüneisen model throughout this work \cite{ziman, HallZhang,pi2024_ZrTe5,zhliu2024}.

\begin{figure*}[htpb]
    \centering
    \includegraphics[width=1\linewidth]{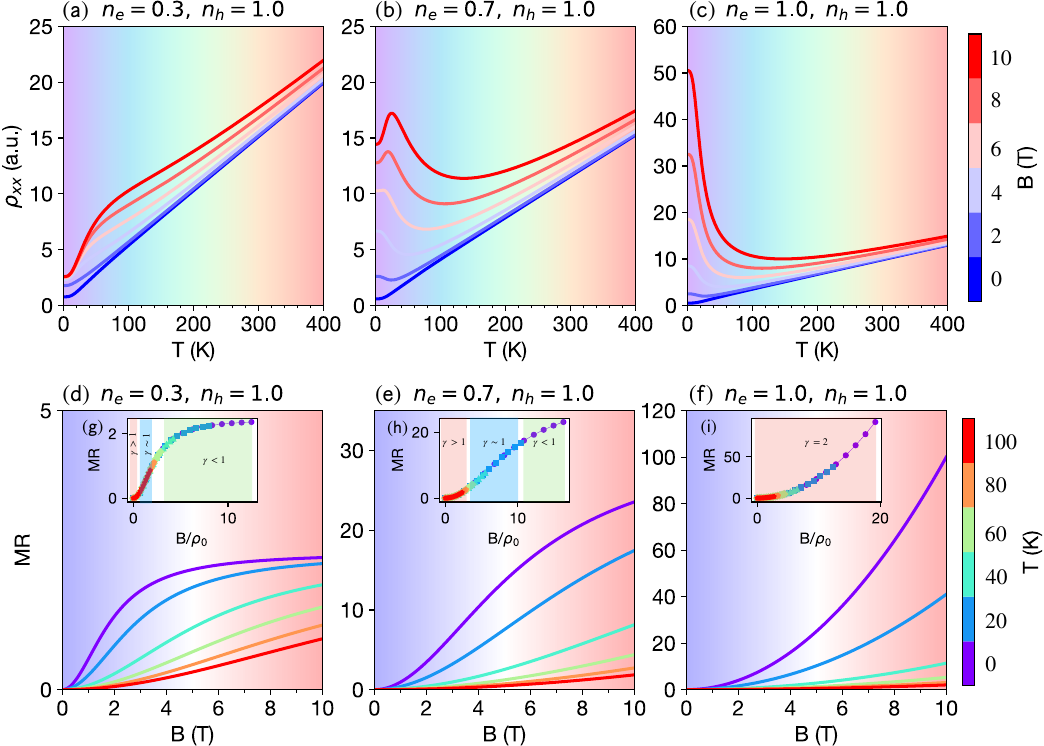}
    \caption{Temperature- and field-dependent resistivity at different exponents $\gamma$. The upper panel displays the temperature-dependent $\rho(T)$, while the lower panel illustrates the field-dependent MR(B). The color of the $\rho$ ($ \rm MR$) curve  matches the color bar corresponding to the magnetic field $B$ (temperature $T$); the background color of the $\rho$ ($\rm MR$) graphic indicates the numerical values of temperature $T$ (field strength $B$), representing areas of low temperature (low field), intermediate temperature (intermediate field), and high temperature (high field). The insets (g), (h) and (i) show the scaled $\MR(B/\rho_0)$ curves at different temperatures.}
    \label{fig2}
\end{figure*}
For the case where $\gamma < 1$, both terms on the right side of Eq. (\ref{equ2}) remain proportional to $\rho_0$. Therefore, the longitudinal resistivity $\rho(T)$, still exhibits a consistent increase with temperature like $\rho_0(T)$. However, the rate of increase in the second term of Eq. (\ref{equ2}) differs from that of the first, resulting in a series of curves that are unevenly spaced, as depicted in Fig. \ref{fig1}(a) for $\gamma = 0.3$. This contrasts with the observations in the $\gamma = 1$ case. 

When the exponent of the power law exceeds one, $\gamma > 1$, the situation becomes complex. The opposite dependence of the two terms in Eq. (\ref{equ2}) on temperature leads to a competition between each other, and thus results in a turn point of the $\rho(T)$ curves. For example, the $\rho(T)$ curves for $B=1$ and $B=0$, as shown in Fig. \ref{fig1}(d) ($\gamma = 2$), exhibit distinctly different behaviors in the low-temperature range. For $B=0$, the resistivity $\rho$ invariably rises with temperature. However, for $B=1$, the resistivity initially declines in the low-temperature range before increasing, a behavior that becomes more pronounced at $B=2$. This physics picture is just the origin of the metal-insulator like behavior which was discussed in Ref.\cite{WTe2Kohler}.

The reason we plot both $\gamma = 1.5$ and $\gamma = 2$ is to observe the increasing prominence of the resistivity reversal trend with temperature as $\gamma$ increases. This observation confirms that such metal-insulator-like characteristics in the $\rho(T)$ curves are more easily detectable experimentally when the charge carriers is close to perfect compensation. Notably, the curves for $\gamma < 1$ and $\gamma > 1$ deviate from the equidistant spacing observed in the $\gamma = 1$ scenario due to the influence of the second term $\frac{B^{\gamma}}{(\alpha\rho_0)^{\gamma-1}}$ in Eq. (\ref{equ2}). For $\gamma < 1$, the spacing between the curves diminishes as the magnetic field increases (due to the diminishing rate of $B^{\gamma}$), but expands as the temperature increases (since the rate of $\rho_0^{1-\gamma}$ accelerates). Conversely, for $\gamma > 1$, the spacing widens with an increase in the magnetic field (as $B^{\gamma}$ grows more rapidly), but narrows as the temperature rises (due to the decreasing rate of $\rho_0^{1-\gamma}$).

The three scenarios discussed above $\gamma < 1$, $\gamma = 1$ and $\gamma > 1$ represent the most basic dependencies of $\rho(B)$. However, in real materials, the power exponent is not a fixed number, but varies with temperature, orientation and strength of the magnetic field. As a result, the behavior of the $\rho(T)$ curve will exhibit more complex characteristics. 

Next, we employ a two-band model to comprehensively analyze the variations of the temperature dependent resistivity for charge
carriers under different degrees of compensation, i.e. (a) $n_e = 0.3$ and $n_h = 1$, (b) $n_e = 0.7$ and $n_h = 1$, (c) $n_e = 1$ and $n_h = 1$. These cases will generate three typical field-dependent $\rho(T)$ curves, which include a monotone rise, the "reentrant metallic" and the "metal-insulator-like transition" behaviors.

{\textbf{A monotone rise behavior}}: Firstly, we discuss the weak compensation case, i.e., $n_e = 0.3$ and $n_h = 1$ for electron and hole's concentration respectively. In Fig.\ref{fig2}(a), a series of $\rho(T)$ curves can be categorized into three temperature zones: low temperature (0-60K), intermediate temperature (80-200K), and high temperature (200K and above). From the inset of Fig.\ref{fig2}(d), it's evident that the region where $\gamma \geq 1$ is quite limited, with most of the curves concentrated in the MR saturation phase (i.e., $\gamma < 1$). These curves can be segmented by temperature and magnetic field. The low-temperature, high-field area corresponds to $\gamma < 1$, as observed in the low-temperature area of Fig.\ref{fig2}(a), where $\rho(T)$ rises monotonically with temperature, colored from pink to red. The progression of these curves resembles the behavior depicted in Fig.\ref{fig1}(a).

As the temperature reaches the intermediate range (80-200K), observations from the inset in Fig.~\ref{fig2}(d) reveal that the $\MR(B)$ approximates a linear relationship (as indicated by light blue shaded region), corresponding to the $\gamma = 1$ case depicted in Fig.~\ref{fig1}(a). During this region, the curves in Fig.~\ref{fig2}(a) transition from light purple to red, maintaining approximately equal spacing that resemble the characteristics observed in Fig. 1(b). As the temperature further escalates into the high temperature zone (200K and above), the MR dependency on the magnetic field surpasses its previous linear trend, as depicted in the light pink region of the inset in Fig.\ref{fig2}(d). This corresponds to the high temperature region where the transition from blue to light purple to red in the curves of Fig.\ref{fig2}(a) illustrates the increasing spacing characteristic of the high temperature area curves shown in Fig.\ref{fig1}(c). Thus, in the case of weak compensation, the behavior of $\rho(T)$ can be comprehensively explained using the patterns and transitions observed in Fig.\ref{fig1}.

\begin{figure*}[htpb]
    \centering
    \includegraphics[width=1\linewidth]{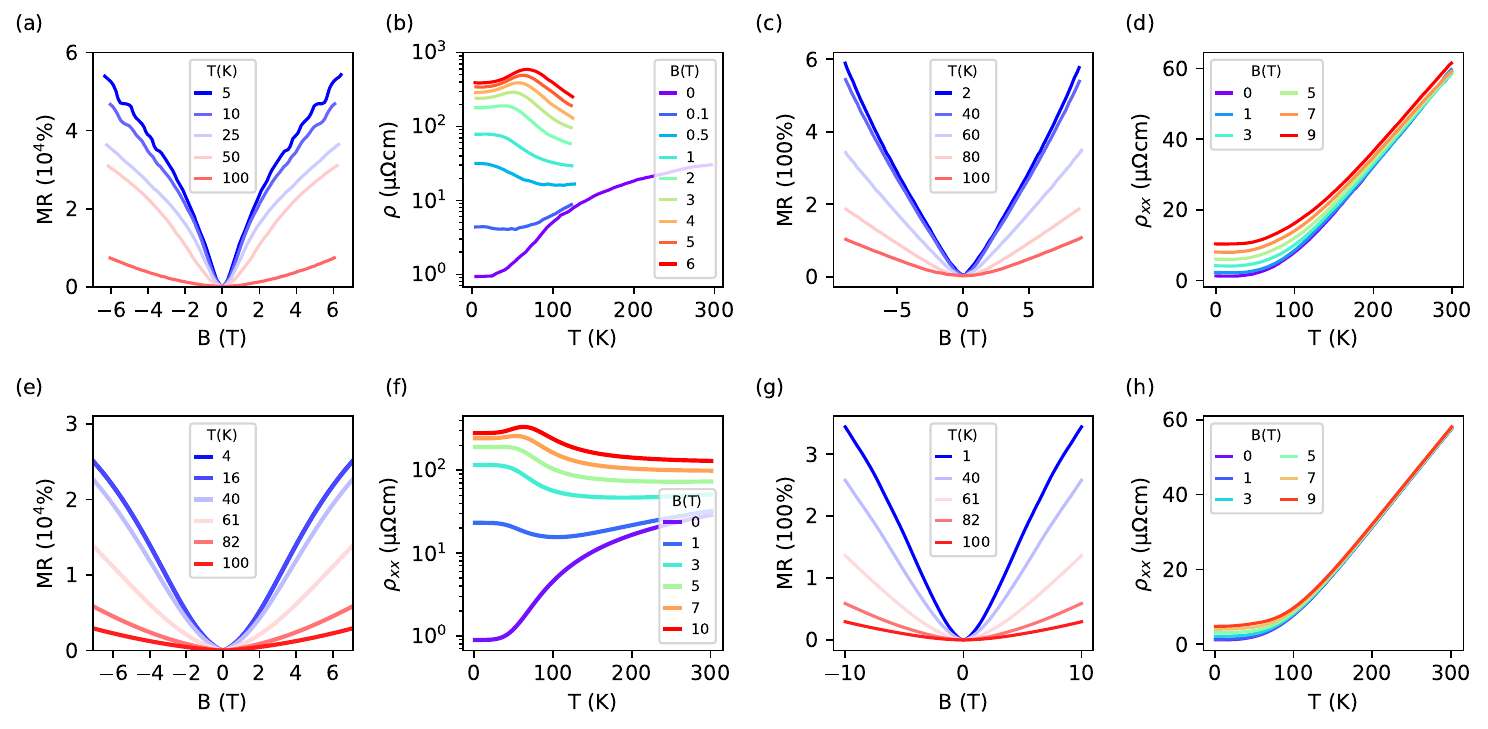}
\caption{Experimentally observed (upper row) and theoretically calculated (bottom row) temperature dependent MR and resistivity of NbP and SiP$_2$. Panels (a) and (b) are reprinted from Ref.~\cite{Sudesh2017} under the Creative Commons Attribution 4.0 International License. (c) and (d) are reprinted from Fig.2(a) and (c) of Ref.~\cite{turnonSiP2}. (e) and (f) show the calculated MR and resistivity of NbP following electron dopping (shifting the Fermi level by 7meV). Panels (g) and (h) are illustrate the calculated MR and resistivity of SiP$_2$ when the magnetic field is applied along the z-direction. }
    \label{fig3}
\end{figure*}
{\textbf{Reentrant behavior}}: Next we discuss the close to compensation case, e.g. $n_e = 0.7$ and $n_h = 1$.  Notably, the area of $ \gamma \geq 1 $ in the inset of Fig.\ref{fig2}(e) is more extensive compared to Fig.\ref{fig2}(d), indicating changes in the behaviors of the intermediate temperature and magnetic field regions. The resistivity behaves similarly with Fig.\ref{fig2}(a) in the low temperature and high field region (i.e., $ \gamma < 1 $): $\rho(T)$ rises with increasing temperature, as drawn by the pink to red curve in Fig.\ref{fig2}(b).

The exponent of ${\rm MR}(B)$ in Fig.\ref{fig2}(e) does not saturate quickly, but involve to $ \gamma >1 $. This introduces a new feature of $\rho(T)$, i.e., $\rho$ decreases with increasing temperature, as illustrated by the light blue to gray curve in the low temperature region of Fig. \ref{fig2} (b). This feature becomes more pronounced as the magnetic field increases because exponent tends to approach $ \gamma =2 $ at slightly higher temperatures, as shown by the pink to red curve in Fig. \ref{fig2} (b) . In the high temperature region, the exponent remains $ \gamma >1 $, thus the $ \rho(T) $ curve replicates the behavior observed in Fig. \ref{fig1}(d) . A similar curve to Fig. \ref{fig2}(b) at $B=10$ was observed in both graphite\cite{reen_grap1, reen_grap2, reen_grap3, reen_grap4} and NbP\cite{reentrantNbP},
, termed "reentrant metallic behavior" by the authors. This is one of the most important conclusions of this work. We begin with the most fundamentals Chambers equation, systematically calculated and elucidated this phenomenon rather than a simple model study. 

{\textbf{Metal-insulator like behavior}}: Finally, let's consider the ideal case where electron hole carriers are perfectly compensated. The ${\rm MR}(B)$ consistently exhibits a quadratic scaling, $\gamma = 2$, regardless of changes in magnetic field and temperature (see inset of Fig.\ref{fig2}(f)). Consequently, $\rho(T)$ in Fig.\ref{fig2}(c) precisely replicates the characteristics observed in Fig.\ref{fig1}(d). In real materials, perfect compensation is rare, but most materials reported to exhibit large MR are nearly perfectly compensated like the case of Fig.\ref{fig1}(c), ensuring that the actual  $\rho(T)$ curve remains close to the ideal case. Therefore experimental observed $\rho(T)$ curves often resemble Fig.\ref{fig2}(c).  

From the analysis above, we understand that the shape of the ${\rm MR}(B)$ curve plays a critical role in shaping the $\rho(T)$ curve. However, the "reentrant-metallic behavior" depicted in Fig.\ref{fig2}(b) is seldom observed in experiments measuring temperature dependent resistivity. The conditions for observing a peak on the temperature-dependent resistivity curves are stringent. Specifically, the ${\rm MR}(B)$ curve must exhibit a $\gamma < 1$ phase to ensure that resistivity increases at low temperatures, and it also requires a $\gamma > 1$ phase to facilitate a decline in resistivity $\rho(T)$ curve  with increasing temperature. Existing experimental data meeting these criteria are scarce. Next we shall compute and analyze the MR and resistivity of two real materials NbP and SiP$_2$. The top panel of Fig.\ref{fig3} displays experimental measurements of NbP and SiP$_2$ from Ref.\cite{Sudesh2017} and Ref.~\cite{turnonSiP2}, respectively, while the bottom panel presents our numerical simulation results. Both materials show very good agreement. Details of the calculations are included in the supplementary materials~\cite{supp}. 

{\textbf{Real material NbP:}} NbP, a typical Weyl semimetal\cite{WSMNbP1, WSMNbP2}, has been observed by several experimental groups to exhibit a peak in the low-temperature $\rho(T)$ curves\cite{kohlerNbP, reentrantNbP, Sudesh2017}, similar to the peak observed in Fig.\ref{fig2}(b). This phenomenon, referred to as "reentrant metallic behavior" \cite{reen_grap1, reentrantNbP}, is depicted in Fig.\ref{fig3} (a)-(b) which show the ${\rm MR}(B)$ and $\rho(T)$ curves derived from Ref.~\cite{Sudesh2017}. At low temperature, the field dependence MR of NbP display a crossover from parabolic to linear. Above 100 K, the MR dependency reverts to parabolic. This behavior satisfies exactly the conditions for "reentrant metallic behavior" in previous discussion on Fig. \ref{fig2}(b). The $\rho(T)$ curve in Fig. \ref{fig3}(b) exhibits a peak just as expected. Our theoretical calculations in Fig.\ref{fig3} (e)-(f) perfectly reproduce these experimental observations, thus validating our theoretical framework. Similar phenomena observed in TaP are also consistent with these findings but are not elaborated on here. We want to point out that the chemical potential of specific sample NbP are critical in observing "reentrant behavior". Our theoretical calculations show that clean samples give rise to the "metal-insulator-like" behavior, while doped ones to the "reentrant behavior"~\cite{supp}.

{\textbf{Real material SiP$_2$:}} SiP$_2$ is a topologically trivial semimetal with highly anistropic Fermi surfaces\cite{supp}. As a result, the scaling of the ${\rm MR}(B)$ curve  depends on the orientation of magnetic field, and hence may exhibit from sub-linear to sub-parabolic characteristics. Due to space constraints, we only analyze the case of $B \parallel $ $z$-axis($\theta=0$) ~\cite{supp}, where the $\rho(B)$ is almost linearly dependent ($\gamma\simeq1.2$) on $B$ at low temperature and quadratically dependent on $B$ at high temperatures as shown in Fig.~\ref{fig3}(c) and (g). In Fig.~\ref{fig3}(d) and (h), the $\rho(T)$ curves of different magnetic fields form a series of monotonically temperature dependence curves. It does not precisely replicate Fig.~\ref{fig1}(b) due to $\gamma \neq 1$ for SiP$_2$, resulting in non-equally spaced curves. The spacing grows larger as the magnetic field increases just like we analyse the curve spacing in Fig.~\ref{fig1}(d) previously. Furthermore, since ${\rm MR}(B)$ deviates only slightly from linearity, there is no characteristic pattern of decline followed by a rise as observed in Fig.~\ref{fig1}(c) and (d). Our calculated results agree well with the experimental data as expected. 

Finally, it is important to emphasize that although our analysis is based on the assumption of Kohler’s rule, the conclusions are universally valid, even in systems beyond Kohler's rule. Kohler’s rule may fail when the relaxation times of multiple carriers differ significantly and lack common factors. We postulate that the electron and hole carriers vary with temperature as shown in Fig.S2\cite{supp}, and accordingly plot the curves of ${\rm MR}(B)$ and $\rho(T)$ for the system in Fig.S2. Despite some discrepancies in detail, it is readily apparent that the overall behavior and trends of ${\rm MR}(B)$ and $\rho(T)$ consistent with scenarios where Kohler’s rule is satisfied. 

%\section{Conclusion}
In this work, we have thoroughly calculated and explained "reentrant-metallic" and "metal-insulator-like" behaviors in the same framework. The $\rho(T)$ curves for different materials can be derived from three fundamental scaling of ${\rm MR}(B)$ as $\gamma<, =, >1$. We then applied our theory to real materials NbP and SiP$_2$ to analyse the MR and resistivity under different magnetic field and temperature. We found very good agreement between our calculations and experimental measurements. This implies that magnetic field can be scaled like temperature when study the resistivity. Therefore the $\rho(T)$ curve cannot simply be used as a metal-insulator phase transition evidence in the presence of a magnetic field.

\begin{acknowledgments}
This work was supported by the National Key R\&D Program of China (Grant No. 2023YFA1607400, 2022YFA1403800), the National Natural Science Foundation of China (Grant No.12274436, 11925408, 11921004), the Science Center of the National Natural Science Foundation of China (Grant No. 12188101), and  H.W. acknowledge support from the New Cornerstone Science Foundation through the XPLORER PRIZE. 
\end{acknowledgments}

\bibliography{refs}

\begin{thebibliography}{48}%
\makeatletter
\providecommand \@ifxundefined [1]{%
 \@ifx{#1\undefined}
}%
\providecommand \@ifnum [1]{%
 \ifnum #1\expandafter \@firstoftwo
 \else \expandafter \@secondoftwo
 \fi
}%
\providecommand \@ifx [1]{%
 \ifx #1\expandafter \@firstoftwo
 \else \expandafter \@secondoftwo
 \fi
}%
\providecommand \natexlab [1]{#1}%
\providecommand \enquote  [1]{``#1''}%
\providecommand \bibnamefont  [1]{#1}%
\providecommand \bibfnamefont [1]{#1}%
\providecommand \citenamefont [1]{#1}%
\providecommand \href@noop [0]{\@secondoftwo}%
\providecommand \href [0]{\begingroup \@sanitize@url \@href}%
\providecommand \@href[1]{\@@startlink{#1}\@@href}%
\providecommand \@@href[1]{\endgroup#1\@@endlink}%
\providecommand \@sanitize@url [0]{\catcode `\\12\catcode `\$12\catcode
  `\&12\catcode `\#12\catcode `\^12\catcode `\_12\catcode `\%12\relax}%
\providecommand \@@startlink[1]{}%
\providecommand \@@endlink[0]{}%
\providecommand \url  [0]{\begingroup\@sanitize@url \@url }%
\providecommand \@url [1]{\endgroup\@href {#1}{\urlprefix }}%
\providecommand \urlprefix  [0]{URL }%
\providecommand \Eprint [0]{\href }%
\providecommand \doibase [0]{http://dx.doi.org/}%
\providecommand \selectlanguage [0]{\@gobble}%
\providecommand \bibinfo  [0]{\@secondoftwo}%
\providecommand \bibfield  [0]{\@secondoftwo}%
\providecommand \translation [1]{[#1]}%
\providecommand \BibitemOpen [0]{}%
\providecommand \bibitemStop [0]{}%
\providecommand \bibitemNoStop [0]{.\EOS\space}%
\providecommand \EOS [0]{\spacefactor3000\relax}%
\providecommand \BibitemShut  [1]{\csname bibitem#1\endcsname}%
\let\auto@bib@innerbib\@empty
%</preamble>
\bibitem [{\citenamefont {Ashcroft}\ and\ \citenamefont
  {Mermin}(1976)}]{Mermin1976}%
  \BibitemOpen
  \bibfield  {author} {\bibinfo {author} {\bibfnamefont {N.~W.}\ \bibnamefont
  {Ashcroft}}\ and\ \bibinfo {author} {\bibfnamefont {N.~D.}\ \bibnamefont
  {Mermin}},\ }\href@noop {} {\emph {\bibinfo {title} {Solid State Physics}}}\
  (\bibinfo  {publisher} {Thomson Learning},\ \bibinfo {year}
  {1976})\BibitemShut {NoStop}%
\bibitem [{\citenamefont {Ziman}(1962)}]{ziman}%
  \BibitemOpen
  \bibfield  {author} {\bibinfo {author} {\bibfnamefont {M.}~\bibnamefont
  {Ziman}},\ }\href@noop {} {\emph {\bibinfo {title} {Electrons and Phonons}}}\
  (\bibinfo  {publisher} {Clarendon Press},\ \bibinfo {address} {Oxford},\
  \bibinfo {year} {1962})\BibitemShut {NoStop}%
\bibitem [{\citenamefont {Kopelevich}\ \emph {et~al.}(1999)\citenamefont
  {Kopelevich}, \citenamefont {Lemanov}, \citenamefont {Moehlecke},\ and\
  \citenamefont {Torres}}]{graphiteMR1}%
  \BibitemOpen
  \bibfield  {author} {\bibinfo {author} {\bibfnamefont {Y.}~\bibnamefont
  {Kopelevich}}, \bibinfo {author} {\bibfnamefont {V.~V.}\ \bibnamefont
  {Lemanov}}, \bibinfo {author} {\bibfnamefont {S.}~\bibnamefont {Moehlecke}},
  \ and\ \bibinfo {author} {\bibfnamefont {J.~H.~S.}\ \bibnamefont {Torres}},\
  }\href {\doibase 10.1134/1.1131135} {\bibfield  {journal} {\bibinfo
  {journal} {Physics of the Solid State}\ }\textbf {\bibinfo {volume} {41}},\
  \bibinfo {pages} {1959–1962} (\bibinfo {year} {1999})}\BibitemShut
  {NoStop}%
\bibitem [{\citenamefont {Kempa}\ \emph {et~al.}(2000)\citenamefont {Kempa},
  \citenamefont {Kopelevich}, \citenamefont {Mrowka}, \citenamefont {Setzer},
  \citenamefont {Torres}, \citenamefont {H\"{o}hne},\ and\ \citenamefont
  {Esquinazi}}]{graphiteMR2}%
  \BibitemOpen
  \bibfield  {author} {\bibinfo {author} {\bibfnamefont {H.}~\bibnamefont
  {Kempa}}, \bibinfo {author} {\bibfnamefont {Y.}~\bibnamefont {Kopelevich}},
  \bibinfo {author} {\bibfnamefont {F.}~\bibnamefont {Mrowka}}, \bibinfo
  {author} {\bibfnamefont {A.}~\bibnamefont {Setzer}}, \bibinfo {author}
  {\bibfnamefont {J.}~\bibnamefont {Torres}}, \bibinfo {author} {\bibfnamefont
  {R.}~\bibnamefont {H\"{o}hne}}, \ and\ \bibinfo {author} {\bibfnamefont
  {P.}~\bibnamefont {Esquinazi}},\ }\href {\doibase
  10.1016/s0038-1098(00)00233-7} {\bibfield  {journal} {\bibinfo  {journal}
  {Solid State Communications}\ }\textbf {\bibinfo {volume} {115}},\ \bibinfo
  {pages} {539–542} (\bibinfo {year} {2000})}\BibitemShut {NoStop}%
\bibitem [{\citenamefont {Kempa}\ \emph {et~al.}(2002)\citenamefont {Kempa},
  \citenamefont {Esquinazi},\ and\ \citenamefont {Kopelevich}}]{graphiteMR3}%
  \BibitemOpen
  \bibfield  {author} {\bibinfo {author} {\bibfnamefont {H.}~\bibnamefont
  {Kempa}}, \bibinfo {author} {\bibfnamefont {P.}~\bibnamefont {Esquinazi}}, \
  and\ \bibinfo {author} {\bibfnamefont {Y.}~\bibnamefont {Kopelevich}},\
  }\href {\doibase 10.1103/PhysRevB.65.241101} {\bibfield  {journal} {\bibinfo
  {journal} {Phys. Rev. B}\ }\textbf {\bibinfo {volume} {65}},\ \bibinfo
  {pages} {241101} (\bibinfo {year} {2002})}\BibitemShut {NoStop}%
\bibitem [{\citenamefont {Kopelevich}\ \emph {et~al.}(2003)\citenamefont
  {Kopelevich}, \citenamefont {Torres}, \citenamefont {da~Silva}, \citenamefont
  {Mrowka}, \citenamefont {Kempa},\ and\ \citenamefont
  {Esquinazi}}]{reen_grap1}%
  \BibitemOpen
  \bibfield  {author} {\bibinfo {author} {\bibfnamefont {Y.}~\bibnamefont
  {Kopelevich}}, \bibinfo {author} {\bibfnamefont {J.~H.~S.}\ \bibnamefont
  {Torres}}, \bibinfo {author} {\bibfnamefont {R.~R.}\ \bibnamefont
  {da~Silva}}, \bibinfo {author} {\bibfnamefont {F.}~\bibnamefont {Mrowka}},
  \bibinfo {author} {\bibfnamefont {H.}~\bibnamefont {Kempa}}, \ and\ \bibinfo
  {author} {\bibfnamefont {P.}~\bibnamefont {Esquinazi}},\ }\href {\doibase
  10.1103/PhysRevLett.90.156402} {\bibfield  {journal} {\bibinfo  {journal}
  {Phys. Rev. Lett.}\ }\textbf {\bibinfo {volume} {90}},\ \bibinfo {pages}
  {156402} (\bibinfo {year} {2003})}\BibitemShut {NoStop}%
\bibitem [{\citenamefont {Tokumoto}\ \emph {et~al.}(2004)\citenamefont
  {Tokumoto}, \citenamefont {Jobiliong}, \citenamefont {Choi}, \citenamefont
  {Oshima},\ and\ \citenamefont {Brooks}}]{graphiteMR4}%
  \BibitemOpen
  \bibfield  {author} {\bibinfo {author} {\bibfnamefont {T.}~\bibnamefont
  {Tokumoto}}, \bibinfo {author} {\bibfnamefont {E.}~\bibnamefont {Jobiliong}},
  \bibinfo {author} {\bibfnamefont {E.}~\bibnamefont {Choi}}, \bibinfo {author}
  {\bibfnamefont {Y.}~\bibnamefont {Oshima}}, \ and\ \bibinfo {author}
  {\bibfnamefont {J.}~\bibnamefont {Brooks}},\ }\href {\doibase
  10.1016/j.ssc.2003.11.037} {\bibfield  {journal} {\bibinfo  {journal} {Solid
  State Communications}\ }\textbf {\bibinfo {volume} {129}},\ \bibinfo {pages}
  {599–604} (\bibinfo {year} {2004})}\BibitemShut {NoStop}%
\bibitem [{\citenamefont {Zhang}\ \emph {et~al.}(2004)\citenamefont {Zhang},
  \citenamefont {Xue},\ and\ \citenamefont {Zhu}}]{graphiteMR5}%
  \BibitemOpen
  \bibfield  {author} {\bibinfo {author} {\bibfnamefont {X.}~\bibnamefont
  {Zhang}}, \bibinfo {author} {\bibfnamefont {Q.}~\bibnamefont {Xue}}, \ and\
  \bibinfo {author} {\bibfnamefont {D.}~\bibnamefont {Zhu}},\ }\href {\doibase
  10.1016/j.physleta.2003.11.050} {\bibfield  {journal} {\bibinfo  {journal}
  {Physics Letters A}\ }\textbf {\bibinfo {volume} {320}},\ \bibinfo {pages}
  {471–477} (\bibinfo {year} {2004})}\BibitemShut {NoStop}%
\bibitem [{\citenamefont {Du}\ \emph {et~al.}(2005)\citenamefont {Du},
  \citenamefont {Tsai}, \citenamefont {Maslov},\ and\ \citenamefont
  {Hebard}}]{grap_scaling}%
  \BibitemOpen
  \bibfield  {author} {\bibinfo {author} {\bibfnamefont {X.}~\bibnamefont
  {Du}}, \bibinfo {author} {\bibfnamefont {S.-W.}\ \bibnamefont {Tsai}},
  \bibinfo {author} {\bibfnamefont {D.~L.}\ \bibnamefont {Maslov}}, \ and\
  \bibinfo {author} {\bibfnamefont {A.~F.}\ \bibnamefont {Hebard}},\ }\href
  {\doibase 10.1103/PhysRevLett.94.166601} {\bibfield  {journal} {\bibinfo
  {journal} {Phys. Rev. Lett.}\ }\textbf {\bibinfo {volume} {94}},\ \bibinfo
  {pages} {166601} (\bibinfo {year} {2005})}\BibitemShut {NoStop}%
\bibitem [{\citenamefont {Kopelevich}\ \emph
  {et~al.}(2006{\natexlab{a}})\citenamefont {Kopelevich}, \citenamefont
  {Pantoja}, \citenamefont {da~Silva},\ and\ \citenamefont
  {Moehlecke}}]{reen_grap2}%
  \BibitemOpen
  \bibfield  {author} {\bibinfo {author} {\bibfnamefont {Y.}~\bibnamefont
  {Kopelevich}}, \bibinfo {author} {\bibfnamefont {J.~C.~M.}\ \bibnamefont
  {Pantoja}}, \bibinfo {author} {\bibfnamefont {R.~R.}\ \bibnamefont
  {da~Silva}}, \ and\ \bibinfo {author} {\bibfnamefont {S.}~\bibnamefont
  {Moehlecke}},\ }\href {\doibase 10.1103/PhysRevB.73.165128} {\bibfield
  {journal} {\bibinfo  {journal} {Phys. Rev. B}\ }\textbf {\bibinfo {volume}
  {73}},\ \bibinfo {pages} {165128} (\bibinfo {year}
  {2006}{\natexlab{a}})}\BibitemShut {NoStop}%
\bibitem [{\citenamefont {Kopelevich}\ \emph
  {et~al.}(2006{\natexlab{b}})\citenamefont {Kopelevich}, \citenamefont
  {{Medina Pantoja}}, \citenamefont {{da Silva}},\ and\ \citenamefont
  {Moehlecke}}]{reen_grap3}%
  \BibitemOpen
  \bibfield  {author} {\bibinfo {author} {\bibfnamefont {Y.}~\bibnamefont
  {Kopelevich}}, \bibinfo {author} {\bibfnamefont {J.}~\bibnamefont {{Medina
  Pantoja}}}, \bibinfo {author} {\bibfnamefont {R.}~\bibnamefont {{da Silva}}},
  \ and\ \bibinfo {author} {\bibfnamefont {S.}~\bibnamefont {Moehlecke}},\
  }\href {\doibase https://doi.org/10.1016/j.aop.2006.04.002} {\bibfield
  {journal} {\bibinfo  {journal} {Annals of Physics}\ }\textbf {\bibinfo
  {volume} {321}},\ \bibinfo {pages} {1575} (\bibinfo {year}
  {2006}{\natexlab{b}})},\ \bibinfo {note} {july 2006 Special
  Issue}\BibitemShut {NoStop}%
\bibitem [{\citenamefont {Konenkova}\ \emph {et~al.}(2008)\citenamefont
  {Konenkova}, \citenamefont {Grundler}, \citenamefont {Morgenstern},\ and\
  \citenamefont {Wiesendanger}}]{reen_grap4}%
  \BibitemOpen
  \bibfield  {author} {\bibinfo {author} {\bibfnamefont {E.~V.}\ \bibnamefont
  {Konenkova}}, \bibinfo {author} {\bibfnamefont {D.}~\bibnamefont {Grundler}},
  \bibinfo {author} {\bibfnamefont {M.}~\bibnamefont {Morgenstern}}, \ and\
  \bibinfo {author} {\bibfnamefont {R.}~\bibnamefont {Wiesendanger}},\ }\href
  {\doibase 10.1134/s1063785008010094} {\bibfield  {journal} {\bibinfo
  {journal} {Technical Physics Letters}\ }\textbf {\bibinfo {volume} {34}},\
  \bibinfo {pages} {30–33} (\bibinfo {year} {2008})}\BibitemShut {NoStop}%
\bibitem [{\citenamefont {Barzola-Quiquia}\ \emph {et~al.}(2019)\citenamefont
  {Barzola-Quiquia}, \citenamefont {Esquinazi}, \citenamefont {Precker},
  \citenamefont {Stiller}, \citenamefont {Zoraghi}, \citenamefont {F\"orster},
  \citenamefont {Herrmannsd\"orfer},\ and\ \citenamefont
  {Coniglio}}]{graphiteMR6}%
  \BibitemOpen
  \bibfield  {author} {\bibinfo {author} {\bibfnamefont {J.}~\bibnamefont
  {Barzola-Quiquia}}, \bibinfo {author} {\bibfnamefont {P.~D.}\ \bibnamefont
  {Esquinazi}}, \bibinfo {author} {\bibfnamefont {C.~E.}\ \bibnamefont
  {Precker}}, \bibinfo {author} {\bibfnamefont {M.}~\bibnamefont {Stiller}},
  \bibinfo {author} {\bibfnamefont {M.}~\bibnamefont {Zoraghi}}, \bibinfo
  {author} {\bibfnamefont {T.}~\bibnamefont {F\"orster}}, \bibinfo {author}
  {\bibfnamefont {T.}~\bibnamefont {Herrmannsd\"orfer}}, \ and\ \bibinfo
  {author} {\bibfnamefont {W.~A.}\ \bibnamefont {Coniglio}},\ }\href {\doibase
  10.1103/PhysRevMaterials.3.054603} {\bibfield  {journal} {\bibinfo  {journal}
  {Phys. Rev. Mater.}\ }\textbf {\bibinfo {volume} {3}},\ \bibinfo {pages}
  {054603} (\bibinfo {year} {2019})}\BibitemShut {NoStop}%
\bibitem [{\citenamefont {Ali}\ \emph {et~al.}(2014)\citenamefont {Ali},
  \citenamefont {Xiong}, \citenamefont {Flynn}, \citenamefont {Tao},
  \citenamefont {Gibson}, \citenamefont {Schoop}, \citenamefont {Liang},
  \citenamefont {Haldolaarachchige}, \citenamefont {Hirschberger},
  \citenamefont {Ong},\ and\ \citenamefont {Cava}}]{MRWTe2nature}%
  \BibitemOpen
  \bibfield  {author} {\bibinfo {author} {\bibfnamefont {M.~N.}\ \bibnamefont
  {Ali}}, \bibinfo {author} {\bibfnamefont {J.}~\bibnamefont {Xiong}}, \bibinfo
  {author} {\bibfnamefont {S.}~\bibnamefont {Flynn}}, \bibinfo {author}
  {\bibfnamefont {J.}~\bibnamefont {Tao}}, \bibinfo {author} {\bibfnamefont
  {Q.~D.}\ \bibnamefont {Gibson}}, \bibinfo {author} {\bibfnamefont {L.~M.}\
  \bibnamefont {Schoop}}, \bibinfo {author} {\bibfnamefont {T.}~\bibnamefont
  {Liang}}, \bibinfo {author} {\bibfnamefont {N.}~\bibnamefont
  {Haldolaarachchige}}, \bibinfo {author} {\bibfnamefont {M.}~\bibnamefont
  {Hirschberger}}, \bibinfo {author} {\bibfnamefont {N.~P.}\ \bibnamefont
  {Ong}}, \ and\ \bibinfo {author} {\bibfnamefont {R.~J.}\ \bibnamefont
  {Cava}},\ }\href {\doibase 10.1038/nature13763} {\bibfield  {journal}
  {\bibinfo  {journal} {Nature}\ }\textbf {\bibinfo {volume} {514}},\ \bibinfo
  {pages} {205} (\bibinfo {year} {2014})}\BibitemShut {NoStop}%
\bibitem [{\citenamefont {Wang}\ \emph {et~al.}(2015)\citenamefont {Wang},
  \citenamefont {Thoutam}, \citenamefont {Xiao}, \citenamefont {Hu},
  \citenamefont {Das}, \citenamefont {Mao}, \citenamefont {Wei}, \citenamefont
  {Divan}, \citenamefont {Luican-Mayer}, \citenamefont {Crabtree},\ and\
  \citenamefont {Kwok}}]{WTe2Kohler}%
  \BibitemOpen
  \bibfield  {author} {\bibinfo {author} {\bibfnamefont {Y.~L.}\ \bibnamefont
  {Wang}}, \bibinfo {author} {\bibfnamefont {L.~R.}\ \bibnamefont {Thoutam}},
  \bibinfo {author} {\bibfnamefont {Z.~L.}\ \bibnamefont {Xiao}}, \bibinfo
  {author} {\bibfnamefont {J.}~\bibnamefont {Hu}}, \bibinfo {author}
  {\bibfnamefont {S.}~\bibnamefont {Das}}, \bibinfo {author} {\bibfnamefont
  {Z.~Q.}\ \bibnamefont {Mao}}, \bibinfo {author} {\bibfnamefont
  {J.}~\bibnamefont {Wei}}, \bibinfo {author} {\bibfnamefont {R.}~\bibnamefont
  {Divan}}, \bibinfo {author} {\bibfnamefont {A.}~\bibnamefont {Luican-Mayer}},
  \bibinfo {author} {\bibfnamefont {G.~W.}\ \bibnamefont {Crabtree}}, \ and\
  \bibinfo {author} {\bibfnamefont {W.~K.}\ \bibnamefont {Kwok}},\ }\href
  {\doibase 10.1103/PhysRevB.92.180402} {\bibfield  {journal} {\bibinfo
  {journal} {Phys. Rev. B}\ }\textbf {\bibinfo {volume} {92}},\ \bibinfo
  {pages} {180402} (\bibinfo {year} {2015})}\BibitemShut {NoStop}%
\bibitem [{\citenamefont {Narayanan}\ \emph {et~al.}(2015)\citenamefont
  {Narayanan}, \citenamefont {Watson}, \citenamefont {Blake}, \citenamefont
  {Bruyant}, \citenamefont {Drigo}, \citenamefont {Chen}, \citenamefont
  {Prabhakaran}, \citenamefont {Yan}, \citenamefont {Felser}, \citenamefont
  {Kong}, \citenamefont {Canfield},\ and\ \citenamefont
  {Coldea}}]{turnonCd3As2}%
  \BibitemOpen
  \bibfield  {author} {\bibinfo {author} {\bibfnamefont {A.}~\bibnamefont
  {Narayanan}}, \bibinfo {author} {\bibfnamefont {M.~D.}\ \bibnamefont
  {Watson}}, \bibinfo {author} {\bibfnamefont {S.~F.}\ \bibnamefont {Blake}},
  \bibinfo {author} {\bibfnamefont {N.}~\bibnamefont {Bruyant}}, \bibinfo
  {author} {\bibfnamefont {L.}~\bibnamefont {Drigo}}, \bibinfo {author}
  {\bibfnamefont {Y.~L.}\ \bibnamefont {Chen}}, \bibinfo {author}
  {\bibfnamefont {D.}~\bibnamefont {Prabhakaran}}, \bibinfo {author}
  {\bibfnamefont {B.}~\bibnamefont {Yan}}, \bibinfo {author} {\bibfnamefont
  {C.}~\bibnamefont {Felser}}, \bibinfo {author} {\bibfnamefont
  {T.}~\bibnamefont {Kong}}, \bibinfo {author} {\bibfnamefont {P.~C.}\
  \bibnamefont {Canfield}}, \ and\ \bibinfo {author} {\bibfnamefont {A.~I.}\
  \bibnamefont {Coldea}},\ }\href {\doibase 10.1103/PhysRevLett.114.117201}
  {\bibfield  {journal} {\bibinfo  {journal} {Phys. Rev. Lett.}\ }\textbf
  {\bibinfo {volume} {114}},\ \bibinfo {pages} {117201} (\bibinfo {year}
  {2015})}\BibitemShut {NoStop}%
\bibitem [{\citenamefont {Han}\ \emph {et~al.}(2017)\citenamefont {Han},
  \citenamefont {Xu}, \citenamefont {Botana}, \citenamefont {Xiao},
  \citenamefont {Wang}, \citenamefont {Yang}, \citenamefont {Chung},
  \citenamefont {Kanatzidis}, \citenamefont {Norman}, \citenamefont
  {Crabtree},\ and\ \citenamefont {Kwok}}]{turnonLaSb}%
  \BibitemOpen
  \bibfield  {author} {\bibinfo {author} {\bibfnamefont {F.}~\bibnamefont
  {Han}}, \bibinfo {author} {\bibfnamefont {J.}~\bibnamefont {Xu}}, \bibinfo
  {author} {\bibfnamefont {A.~S.}\ \bibnamefont {Botana}}, \bibinfo {author}
  {\bibfnamefont {Z.~L.}\ \bibnamefont {Xiao}}, \bibinfo {author}
  {\bibfnamefont {Y.~L.}\ \bibnamefont {Wang}}, \bibinfo {author}
  {\bibfnamefont {W.~G.}\ \bibnamefont {Yang}}, \bibinfo {author}
  {\bibfnamefont {D.~Y.}\ \bibnamefont {Chung}}, \bibinfo {author}
  {\bibfnamefont {M.~G.}\ \bibnamefont {Kanatzidis}}, \bibinfo {author}
  {\bibfnamefont {M.~R.}\ \bibnamefont {Norman}}, \bibinfo {author}
  {\bibfnamefont {G.~W.}\ \bibnamefont {Crabtree}}, \ and\ \bibinfo {author}
  {\bibfnamefont {W.~K.}\ \bibnamefont {Kwok}},\ }\href {\doibase
  10.1103/PhysRevB.96.125112} {\bibfield  {journal} {\bibinfo  {journal} {Phys.
  Rev. B}\ }\textbf {\bibinfo {volume} {96}},\ \bibinfo {pages} {125112}
  (\bibinfo {year} {2017})}\BibitemShut {NoStop}%
\bibitem [{\citenamefont {Xu}\ \emph {et~al.}(2017{\natexlab{a}})\citenamefont
  {Xu}, \citenamefont {Bugaris}, \citenamefont {Xiao}, \citenamefont {Wang},
  \citenamefont {Chung}, \citenamefont {Kanatzidis},\ and\ \citenamefont
  {Kwok}}]{turnonNbP}%
  \BibitemOpen
  \bibfield  {author} {\bibinfo {author} {\bibfnamefont {J.}~\bibnamefont
  {Xu}}, \bibinfo {author} {\bibfnamefont {D.~E.}\ \bibnamefont {Bugaris}},
  \bibinfo {author} {\bibfnamefont {Z.~L.}\ \bibnamefont {Xiao}}, \bibinfo
  {author} {\bibfnamefont {Y.~L.}\ \bibnamefont {Wang}}, \bibinfo {author}
  {\bibfnamefont {D.~Y.}\ \bibnamefont {Chung}}, \bibinfo {author}
  {\bibfnamefont {M.~G.}\ \bibnamefont {Kanatzidis}}, \ and\ \bibinfo {author}
  {\bibfnamefont {W.~K.}\ \bibnamefont {Kwok}},\ }\href {\doibase
  10.1103/PhysRevB.96.115152} {\bibfield  {journal} {\bibinfo  {journal} {Phys.
  Rev. B}\ }\textbf {\bibinfo {volume} {96}},\ \bibinfo {pages} {115152}
  (\bibinfo {year} {2017}{\natexlab{a}})}\BibitemShut {NoStop}%
\bibitem [{\citenamefont {Pei}\ \emph {et~al.}(2017)\citenamefont {Pei},
  \citenamefont {Meng}, \citenamefont {Luo}, \citenamefont {Lv}, \citenamefont
  {Chen}, \citenamefont {Lu}, \citenamefont {Han}, \citenamefont {Tong},
  \citenamefont {Song}, \citenamefont {Hou}, \citenamefont {Lu},\ and\
  \citenamefont {Sun}}]{turnonMoTe2}%
  \BibitemOpen
  \bibfield  {author} {\bibinfo {author} {\bibfnamefont {Q.~L.}\ \bibnamefont
  {Pei}}, \bibinfo {author} {\bibfnamefont {W.~J.}\ \bibnamefont {Meng}},
  \bibinfo {author} {\bibfnamefont {X.}~\bibnamefont {Luo}}, \bibinfo {author}
  {\bibfnamefont {H.~Y.}\ \bibnamefont {Lv}}, \bibinfo {author} {\bibfnamefont
  {F.~C.}\ \bibnamefont {Chen}}, \bibinfo {author} {\bibfnamefont {W.~J.}\
  \bibnamefont {Lu}}, \bibinfo {author} {\bibfnamefont {Y.~Y.}\ \bibnamefont
  {Han}}, \bibinfo {author} {\bibfnamefont {P.}~\bibnamefont {Tong}}, \bibinfo
  {author} {\bibfnamefont {W.~H.}\ \bibnamefont {Song}}, \bibinfo {author}
  {\bibfnamefont {Y.~B.}\ \bibnamefont {Hou}}, \bibinfo {author} {\bibfnamefont
  {Q.~Y.}\ \bibnamefont {Lu}}, \ and\ \bibinfo {author} {\bibfnamefont {Y.~P.}\
  \bibnamefont {Sun}},\ }\href {\doibase 10.1103/PhysRevB.96.075132} {\bibfield
   {journal} {\bibinfo  {journal} {Phys. Rev. B}\ }\textbf {\bibinfo {volume}
  {96}},\ \bibinfo {pages} {075132} (\bibinfo {year} {2017})}\BibitemShut
  {NoStop}%
\bibitem [{\citenamefont {Sun}\ \emph {et~al.}(2016)\citenamefont {Sun},
  \citenamefont {Wang}, \citenamefont {Guo}, \citenamefont {Liu},\ and\
  \citenamefont {Lei}}]{turnonLaBi}%
  \BibitemOpen
  \bibfield  {author} {\bibinfo {author} {\bibfnamefont {S.}~\bibnamefont
  {Sun}}, \bibinfo {author} {\bibfnamefont {Q.}~\bibnamefont {Wang}}, \bibinfo
  {author} {\bibfnamefont {P.-J.}\ \bibnamefont {Guo}}, \bibinfo {author}
  {\bibfnamefont {K.}~\bibnamefont {Liu}}, \ and\ \bibinfo {author}
  {\bibfnamefont {H.}~\bibnamefont {Lei}},\ }\href {\doibase
  10.1088/1367-2630/18/8/082002} {\bibfield  {journal} {\bibinfo  {journal}
  {New Journal of Physics}\ }\textbf {\bibinfo {volume} {18}},\ \bibinfo
  {pages} {082002} (\bibinfo {year} {2016})}\BibitemShut {NoStop}%
\bibitem [{\citenamefont {Du}\ \emph {et~al.}(2018)\citenamefont {Du},
  \citenamefont {Lou}, \citenamefont {Zhang}, \citenamefont {Zhou},
  \citenamefont {Xu}, \citenamefont {Chen}, \citenamefont {Tang}, \citenamefont
  {Chen}, \citenamefont {Chen}, \citenamefont {Zhu}, \citenamefont {Wang},
  \citenamefont {Yang}, \citenamefont {Wu}, \citenamefont {Yazyev},\ and\
  \citenamefont {Fang}}]{turnonaWP2}%
  \BibitemOpen
  \bibfield  {author} {\bibinfo {author} {\bibfnamefont {J.}~\bibnamefont
  {Du}}, \bibinfo {author} {\bibfnamefont {Z.}~\bibnamefont {Lou}}, \bibinfo
  {author} {\bibfnamefont {S.}~\bibnamefont {Zhang}}, \bibinfo {author}
  {\bibfnamefont {Y.}~\bibnamefont {Zhou}}, \bibinfo {author} {\bibfnamefont
  {B.}~\bibnamefont {Xu}}, \bibinfo {author} {\bibfnamefont {Q.}~\bibnamefont
  {Chen}}, \bibinfo {author} {\bibfnamefont {Y.}~\bibnamefont {Tang}}, \bibinfo
  {author} {\bibfnamefont {S.}~\bibnamefont {Chen}}, \bibinfo {author}
  {\bibfnamefont {H.}~\bibnamefont {Chen}}, \bibinfo {author} {\bibfnamefont
  {Q.}~\bibnamefont {Zhu}}, \bibinfo {author} {\bibfnamefont {H.}~\bibnamefont
  {Wang}}, \bibinfo {author} {\bibfnamefont {J.}~\bibnamefont {Yang}}, \bibinfo
  {author} {\bibfnamefont {Q.}~\bibnamefont {Wu}}, \bibinfo {author}
  {\bibfnamefont {O.~V.}\ \bibnamefont {Yazyev}}, \ and\ \bibinfo {author}
  {\bibfnamefont {M.}~\bibnamefont {Fang}},\ }\href {\doibase
  10.1103/PhysRevB.97.245101} {\bibfield  {journal} {\bibinfo  {journal} {Phys.
  Rev. B}\ }\textbf {\bibinfo {volume} {97}},\ \bibinfo {pages} {245101}
  (\bibinfo {year} {2018})}\BibitemShut {NoStop}%
\bibitem [{\citenamefont {Kumar}\ \emph {et~al.}(2017)\citenamefont {Kumar},
  \citenamefont {Sun}, \citenamefont {Xu}, \citenamefont {Manna}, \citenamefont
  {Yao}, \citenamefont {S{\"{u}}ss}, \citenamefont {Leermakers}, \citenamefont
  {Young}, \citenamefont {F{\"{o}}rster}, \citenamefont {Schmidt},
  \citenamefont {Borrmann}, \citenamefont {Yan}, \citenamefont {Zeitler},
  \citenamefont {Shi}, \citenamefont {Felser},\ and\ \citenamefont
  {Shekhar}}]{turnonbWP2}%
  \BibitemOpen
  \bibfield  {author} {\bibinfo {author} {\bibfnamefont {N.}~\bibnamefont
  {Kumar}}, \bibinfo {author} {\bibfnamefont {Y.}~\bibnamefont {Sun}}, \bibinfo
  {author} {\bibfnamefont {N.}~\bibnamefont {Xu}}, \bibinfo {author}
  {\bibfnamefont {K.}~\bibnamefont {Manna}}, \bibinfo {author} {\bibfnamefont
  {M.}~\bibnamefont {Yao}}, \bibinfo {author} {\bibfnamefont {V.}~\bibnamefont
  {S{\"{u}}ss}}, \bibinfo {author} {\bibfnamefont {I.}~\bibnamefont
  {Leermakers}}, \bibinfo {author} {\bibfnamefont {O.}~\bibnamefont {Young}},
  \bibinfo {author} {\bibfnamefont {T.}~\bibnamefont {F{\"{o}}rster}}, \bibinfo
  {author} {\bibfnamefont {M.}~\bibnamefont {Schmidt}}, \bibinfo {author}
  {\bibfnamefont {H.}~\bibnamefont {Borrmann}}, \bibinfo {author}
  {\bibfnamefont {B.}~\bibnamefont {Yan}}, \bibinfo {author} {\bibfnamefont
  {U.}~\bibnamefont {Zeitler}}, \bibinfo {author} {\bibfnamefont
  {M.}~\bibnamefont {Shi}}, \bibinfo {author} {\bibfnamefont {C.}~\bibnamefont
  {Felser}}, \ and\ \bibinfo {author} {\bibfnamefont {C.}~\bibnamefont
  {Shekhar}},\ }\href {\doibase 10.1038/s41467-017-01758-z} {\bibfield
  {journal} {\bibinfo  {journal} {Nature Communications}\ }\textbf {\bibinfo
  {volume} {8}},\ \bibinfo {pages} {1642} (\bibinfo {year} {2017})}\BibitemShut
  {NoStop}%
\bibitem [{\citenamefont {Chen}\ \emph {et~al.}(2020)\citenamefont {Chen},
  \citenamefont {Lou}, \citenamefont {Zhang}, \citenamefont {Xu}, \citenamefont
  {Zhou}, \citenamefont {Chen}, \citenamefont {Chen}, \citenamefont {Du},
  \citenamefont {Wang}, \citenamefont {Yang}, \citenamefont {Wu}, \citenamefont
  {Yazyev},\ and\ \citenamefont {Fang}}]{turnonMoO2}%
  \BibitemOpen
  \bibfield  {author} {\bibinfo {author} {\bibfnamefont {Q.}~\bibnamefont
  {Chen}}, \bibinfo {author} {\bibfnamefont {Z.}~\bibnamefont {Lou}}, \bibinfo
  {author} {\bibfnamefont {S.}~\bibnamefont {Zhang}}, \bibinfo {author}
  {\bibfnamefont {B.}~\bibnamefont {Xu}}, \bibinfo {author} {\bibfnamefont
  {Y.}~\bibnamefont {Zhou}}, \bibinfo {author} {\bibfnamefont {H.}~\bibnamefont
  {Chen}}, \bibinfo {author} {\bibfnamefont {S.}~\bibnamefont {Chen}}, \bibinfo
  {author} {\bibfnamefont {J.}~\bibnamefont {Du}}, \bibinfo {author}
  {\bibfnamefont {H.}~\bibnamefont {Wang}}, \bibinfo {author} {\bibfnamefont
  {J.}~\bibnamefont {Yang}}, \bibinfo {author} {\bibfnamefont {Q.}~\bibnamefont
  {Wu}}, \bibinfo {author} {\bibfnamefont {O.~V.}\ \bibnamefont {Yazyev}}, \
  and\ \bibinfo {author} {\bibfnamefont {M.}~\bibnamefont {Fang}},\ }\href
  {\doibase 10.1103/PhysRevB.102.165133} {\bibfield  {journal} {\bibinfo
  {journal} {Phys. Rev. B}\ }\textbf {\bibinfo {volume} {102}},\ \bibinfo
  {pages} {165133} (\bibinfo {year} {2020})}\BibitemShut {NoStop}%
\bibitem [{\citenamefont {Zhou}\ \emph {et~al.}(2020)\citenamefont {Zhou},
  \citenamefont {Lou}, \citenamefont {Zhang}, \citenamefont {Chen},
  \citenamefont {Chen}, \citenamefont {Xu}, \citenamefont {Du}, \citenamefont
  {Yang}, \citenamefont {Wang}, \citenamefont {Xi}, \citenamefont {Pi},
  \citenamefont {Wu}, \citenamefont {Yazyev},\ and\ \citenamefont
  {Fang}}]{turnonSiP2}%
  \BibitemOpen
  \bibfield  {author} {\bibinfo {author} {\bibfnamefont {Y.}~\bibnamefont
  {Zhou}}, \bibinfo {author} {\bibfnamefont {Z.}~\bibnamefont {Lou}}, \bibinfo
  {author} {\bibfnamefont {S.}~\bibnamefont {Zhang}}, \bibinfo {author}
  {\bibfnamefont {H.}~\bibnamefont {Chen}}, \bibinfo {author} {\bibfnamefont
  {Q.}~\bibnamefont {Chen}}, \bibinfo {author} {\bibfnamefont {B.}~\bibnamefont
  {Xu}}, \bibinfo {author} {\bibfnamefont {J.}~\bibnamefont {Du}}, \bibinfo
  {author} {\bibfnamefont {J.}~\bibnamefont {Yang}}, \bibinfo {author}
  {\bibfnamefont {H.}~\bibnamefont {Wang}}, \bibinfo {author} {\bibfnamefont
  {C.}~\bibnamefont {Xi}}, \bibinfo {author} {\bibfnamefont {L.}~\bibnamefont
  {Pi}}, \bibinfo {author} {\bibfnamefont {Q.}~\bibnamefont {Wu}}, \bibinfo
  {author} {\bibfnamefont {O.~V.}\ \bibnamefont {Yazyev}}, \ and\ \bibinfo
  {author} {\bibfnamefont {M.}~\bibnamefont {Fang}},\ }\href {\doibase
  10.1103/PhysRevB.102.115145} {\bibfield  {journal} {\bibinfo  {journal}
  {Phys. Rev. B}\ }\textbf {\bibinfo {volume} {102}},\ \bibinfo {pages}
  {115145} (\bibinfo {year} {2020})}\BibitemShut {NoStop}%
\bibitem [{\citenamefont {Chen}\ \emph {et~al.}(2021)\citenamefont {Chen},
  \citenamefont {Lou}, \citenamefont {Zhang}, \citenamefont {Zhou},
  \citenamefont {Xu}, \citenamefont {Chen}, \citenamefont {Chen}, \citenamefont
  {Du}, \citenamefont {Wang}, \citenamefont {Yang}, \citenamefont {Wu},
  \citenamefont {Yazyev},\ and\ \citenamefont {Fang}}]{turnonReO3}%
  \BibitemOpen
  \bibfield  {author} {\bibinfo {author} {\bibfnamefont {Q.}~\bibnamefont
  {Chen}}, \bibinfo {author} {\bibfnamefont {Z.}~\bibnamefont {Lou}}, \bibinfo
  {author} {\bibfnamefont {S.}~\bibnamefont {Zhang}}, \bibinfo {author}
  {\bibfnamefont {Y.}~\bibnamefont {Zhou}}, \bibinfo {author} {\bibfnamefont
  {B.}~\bibnamefont {Xu}}, \bibinfo {author} {\bibfnamefont {H.}~\bibnamefont
  {Chen}}, \bibinfo {author} {\bibfnamefont {S.}~\bibnamefont {Chen}}, \bibinfo
  {author} {\bibfnamefont {J.}~\bibnamefont {Du}}, \bibinfo {author}
  {\bibfnamefont {H.}~\bibnamefont {Wang}}, \bibinfo {author} {\bibfnamefont
  {J.}~\bibnamefont {Yang}}, \bibinfo {author} {\bibfnamefont {Q.}~\bibnamefont
  {Wu}}, \bibinfo {author} {\bibfnamefont {O.~V.}\ \bibnamefont {Yazyev}}, \
  and\ \bibinfo {author} {\bibfnamefont {M.}~\bibnamefont {Fang}},\ }\href
  {\doibase 10.1103/PhysRevB.104.115104} {\bibfield  {journal} {\bibinfo
  {journal} {Phys. Rev. B}\ }\textbf {\bibinfo {volume} {104}},\ \bibinfo
  {pages} {115104} (\bibinfo {year} {2021})}\BibitemShut {NoStop}%
\bibitem [{\citenamefont {Saleheen}\ \emph {et~al.}(2020)\citenamefont
  {Saleheen}, \citenamefont {Chapai}, \citenamefont {Xing}, \citenamefont
  {Nepal}, \citenamefont {Gong}, \citenamefont {Gui}, \citenamefont {Xie},
  \citenamefont {Young}, \citenamefont {Plummer},\ and\ \citenamefont
  {Jin}}]{turnonTaSe3}%
  \BibitemOpen
  \bibfield  {author} {\bibinfo {author} {\bibfnamefont {A.~I.~U.}\
  \bibnamefont {Saleheen}}, \bibinfo {author} {\bibfnamefont {R.}~\bibnamefont
  {Chapai}}, \bibinfo {author} {\bibfnamefont {L.}~\bibnamefont {Xing}},
  \bibinfo {author} {\bibfnamefont {R.}~\bibnamefont {Nepal}}, \bibinfo
  {author} {\bibfnamefont {D.}~\bibnamefont {Gong}}, \bibinfo {author}
  {\bibfnamefont {X.}~\bibnamefont {Gui}}, \bibinfo {author} {\bibfnamefont
  {W.}~\bibnamefont {Xie}}, \bibinfo {author} {\bibfnamefont {D.~P.}\
  \bibnamefont {Young}}, \bibinfo {author} {\bibfnamefont {E.~W.}\ \bibnamefont
  {Plummer}}, \ and\ \bibinfo {author} {\bibfnamefont {R.}~\bibnamefont
  {Jin}},\ }\href {\doibase 10.1038/s41535-020-00257-7} {\bibfield  {journal}
  {\bibinfo  {journal} {npj Quantum Materials}\ }\textbf {\bibinfo {volume}
  {5}},\ \bibinfo {pages} {53} (\bibinfo {year} {2020})}\BibitemShut {NoStop}%
\bibitem [{\citenamefont {Dan}\ \emph {et~al.}(2023)\citenamefont {Dan},
  \citenamefont {Kargeti}, \citenamefont {Sahoo}, \citenamefont {Dan},
  \citenamefont {Pal}, \citenamefont {Verma}, \citenamefont {Chakravarty},
  \citenamefont {Panda},\ and\ \citenamefont {Patil}}]{turnonInBi}%
  \BibitemOpen
  \bibfield  {author} {\bibinfo {author} {\bibfnamefont {S.}~\bibnamefont
  {Dan}}, \bibinfo {author} {\bibfnamefont {K.}~\bibnamefont {Kargeti}},
  \bibinfo {author} {\bibfnamefont {R.~C.}\ \bibnamefont {Sahoo}}, \bibinfo
  {author} {\bibfnamefont {S.}~\bibnamefont {Dan}}, \bibinfo {author}
  {\bibfnamefont {D.}~\bibnamefont {Pal}}, \bibinfo {author} {\bibfnamefont
  {S.}~\bibnamefont {Verma}}, \bibinfo {author} {\bibfnamefont
  {S.}~\bibnamefont {Chakravarty}}, \bibinfo {author} {\bibfnamefont {S.~K.}\
  \bibnamefont {Panda}}, \ and\ \bibinfo {author} {\bibfnamefont
  {S.}~\bibnamefont {Patil}},\ }\href {\doibase 10.1103/PhysRevB.107.205111}
  {\bibfield  {journal} {\bibinfo  {journal} {Phys. Rev. B}\ }\textbf {\bibinfo
  {volume} {107}},\ \bibinfo {pages} {205111} (\bibinfo {year}
  {2023})}\BibitemShut {NoStop}%
\bibitem [{\citenamefont {Dwari}\ \emph {et~al.}(2023)\citenamefont {Dwari},
  \citenamefont {Sasmal}, \citenamefont {Dan}, \citenamefont {Maity},
  \citenamefont {Saini}, \citenamefont {Kulkarni}, \citenamefont {Banik},
  \citenamefont {Verma}, \citenamefont {Singh},\ and\ \citenamefont
  {Thamizhavel}}]{turnonGdBi}%
  \BibitemOpen
  \bibfield  {author} {\bibinfo {author} {\bibfnamefont {G.}~\bibnamefont
  {Dwari}}, \bibinfo {author} {\bibfnamefont {S.}~\bibnamefont {Sasmal}},
  \bibinfo {author} {\bibfnamefont {S.}~\bibnamefont {Dan}}, \bibinfo {author}
  {\bibfnamefont {B.}~\bibnamefont {Maity}}, \bibinfo {author} {\bibfnamefont
  {V.}~\bibnamefont {Saini}}, \bibinfo {author} {\bibfnamefont
  {R.}~\bibnamefont {Kulkarni}}, \bibinfo {author} {\bibfnamefont
  {S.}~\bibnamefont {Banik}}, \bibinfo {author} {\bibfnamefont
  {R.}~\bibnamefont {Verma}}, \bibinfo {author} {\bibfnamefont
  {B.}~\bibnamefont {Singh}}, \ and\ \bibinfo {author} {\bibfnamefont
  {A.}~\bibnamefont {Thamizhavel}},\ }\href {\doibase
  10.1103/PhysRevB.107.235117} {\bibfield  {journal} {\bibinfo  {journal}
  {Phys. Rev. B}\ }\textbf {\bibinfo {volume} {107}},\ \bibinfo {pages}
  {235117} (\bibinfo {year} {2023})}\BibitemShut {NoStop}%
\bibitem [{\citenamefont {Singh}\ \emph {et~al.}(2022)\citenamefont {Singh},
  \citenamefont {Sasmal}, \citenamefont {Iyer}, \citenamefont {Thamizhavel},\
  and\ \citenamefont {Maiti}}]{turnonWTe2Ni}%
  \BibitemOpen
  \bibfield  {author} {\bibinfo {author} {\bibfnamefont {A.}~\bibnamefont
  {Singh}}, \bibinfo {author} {\bibfnamefont {S.}~\bibnamefont {Sasmal}},
  \bibinfo {author} {\bibfnamefont {K.~K.}\ \bibnamefont {Iyer}}, \bibinfo
  {author} {\bibfnamefont {A.}~\bibnamefont {Thamizhavel}}, \ and\ \bibinfo
  {author} {\bibfnamefont {K.}~\bibnamefont {Maiti}},\ }\href {\doibase
  10.1103/PhysRevMaterials.6.124202} {\bibfield  {journal} {\bibinfo  {journal}
  {Phys. Rev. Mater.}\ }\textbf {\bibinfo {volume} {6}},\ \bibinfo {pages}
  {124202} (\bibinfo {year} {2022})}\BibitemShut {NoStop}%
\bibitem [{\citenamefont {Zhang}\ \emph {et~al.}(2015)\citenamefont {Zhang},
  \citenamefont {Guo}, \citenamefont {Lu}, \citenamefont {Zhang}, \citenamefont
  {Yuan}, \citenamefont {Lin}, \citenamefont {Wang},\ and\ \citenamefont
  {Jia}}]{turnonTaP}%
  \BibitemOpen
  \bibfield  {author} {\bibinfo {author} {\bibfnamefont {C.}~\bibnamefont
  {Zhang}}, \bibinfo {author} {\bibfnamefont {C.}~\bibnamefont {Guo}}, \bibinfo
  {author} {\bibfnamefont {H.}~\bibnamefont {Lu}}, \bibinfo {author}
  {\bibfnamefont {X.}~\bibnamefont {Zhang}}, \bibinfo {author} {\bibfnamefont
  {Z.}~\bibnamefont {Yuan}}, \bibinfo {author} {\bibfnamefont {Z.}~\bibnamefont
  {Lin}}, \bibinfo {author} {\bibfnamefont {J.}~\bibnamefont {Wang}}, \ and\
  \bibinfo {author} {\bibfnamefont {S.}~\bibnamefont {Jia}},\ }\href {\doibase
  10.1103/PhysRevB.92.041203} {\bibfield  {journal} {\bibinfo  {journal} {Phys.
  Rev. B}\ }\textbf {\bibinfo {volume} {92}},\ \bibinfo {pages} {041203}
  (\bibinfo {year} {2015})}\BibitemShut {NoStop}%
\bibitem [{\citenamefont {Laha}\ \emph {et~al.}(2020)\citenamefont {Laha},
  \citenamefont {Mardanya}, \citenamefont {Singh}, \citenamefont {Lin},
  \citenamefont {Bansil}, \citenamefont {Agarwal},\ and\ \citenamefont
  {Hossain}}]{turnonCaCdSn}%
  \BibitemOpen
  \bibfield  {author} {\bibinfo {author} {\bibfnamefont {A.}~\bibnamefont
  {Laha}}, \bibinfo {author} {\bibfnamefont {S.}~\bibnamefont {Mardanya}},
  \bibinfo {author} {\bibfnamefont {B.}~\bibnamefont {Singh}}, \bibinfo
  {author} {\bibfnamefont {H.}~\bibnamefont {Lin}}, \bibinfo {author}
  {\bibfnamefont {A.}~\bibnamefont {Bansil}}, \bibinfo {author} {\bibfnamefont
  {A.}~\bibnamefont {Agarwal}}, \ and\ \bibinfo {author} {\bibfnamefont
  {Z.}~\bibnamefont {Hossain}},\ }\href {\doibase 10.1103/PhysRevB.102.035164}
  {\bibfield  {journal} {\bibinfo  {journal} {Phys. Rev. B}\ }\textbf {\bibinfo
  {volume} {102}},\ \bibinfo {pages} {035164} (\bibinfo {year}
  {2020})}\BibitemShut {NoStop}%
\bibitem [{\citenamefont {Xu}\ \emph {et~al.}(2017{\natexlab{b}})\citenamefont
  {Xu}, \citenamefont {Bugaris}, \citenamefont {Xiao}, \citenamefont {Wang},
  \citenamefont {Chung}, \citenamefont {Kanatzidis},\ and\ \citenamefont
  {Kwok}}]{reentrantNbP}%
  \BibitemOpen
  \bibfield  {author} {\bibinfo {author} {\bibfnamefont {J.}~\bibnamefont
  {Xu}}, \bibinfo {author} {\bibfnamefont {D.~E.}\ \bibnamefont {Bugaris}},
  \bibinfo {author} {\bibfnamefont {Z.~L.}\ \bibnamefont {Xiao}}, \bibinfo
  {author} {\bibfnamefont {Y.~L.}\ \bibnamefont {Wang}}, \bibinfo {author}
  {\bibfnamefont {D.~Y.}\ \bibnamefont {Chung}}, \bibinfo {author}
  {\bibfnamefont {M.~G.}\ \bibnamefont {Kanatzidis}}, \ and\ \bibinfo {author}
  {\bibfnamefont {W.~K.}\ \bibnamefont {Kwok}},\ }\href {\doibase
  10.1103/PhysRevB.96.115152} {\bibfield  {journal} {\bibinfo  {journal} {Phys.
  Rev. B}\ }\textbf {\bibinfo {volume} {96}},\ \bibinfo {pages} {115152}
  (\bibinfo {year} {2017}{\natexlab{b}})}\BibitemShut {NoStop}%
\bibitem [{\citenamefont {Kresse}\ and\ \citenamefont
  {Furthm\"uller}(1996)}]{Vasp1}%
  \BibitemOpen
  \bibfield  {author} {\bibinfo {author} {\bibfnamefont {G.}~\bibnamefont
  {Kresse}}\ and\ \bibinfo {author} {\bibfnamefont {J.}~\bibnamefont
  {Furthm\"uller}},\ }\href {\doibase 10.1103/PhysRevB.54.11169} {\bibfield
  {journal} {\bibinfo  {journal} {Phys. Rev. B}\ }\textbf {\bibinfo {volume}
  {54}},\ \bibinfo {pages} {11169} (\bibinfo {year} {1996})}\BibitemShut
  {NoStop}%
\bibitem [{\citenamefont {Kresse}\ and\ \citenamefont {Joubert}(1999)}]{Vasp2}%
  \BibitemOpen
  \bibfield  {author} {\bibinfo {author} {\bibfnamefont {G.}~\bibnamefont
  {Kresse}}\ and\ \bibinfo {author} {\bibfnamefont {D.}~\bibnamefont
  {Joubert}},\ }\href {\doibase 10.1103/PhysRevB.59.1758} {\bibfield  {journal}
  {\bibinfo  {journal} {Phys. Rev. B}\ }\textbf {\bibinfo {volume} {59}},\
  \bibinfo {pages} {1758} (\bibinfo {year} {1999})}\BibitemShut {NoStop}%
\bibitem [{\citenamefont {Mostofi}\ \emph {et~al.}(2014)\citenamefont
  {Mostofi}, \citenamefont {Yates}, \citenamefont {Pizzi}, \citenamefont {Lee},
  \citenamefont {Souza}, \citenamefont {Vanderbilt},\ and\ \citenamefont
  {Marzari}}]{NMwannier90}%
  \BibitemOpen
  \bibfield  {author} {\bibinfo {author} {\bibfnamefont {A.~A.}\ \bibnamefont
  {Mostofi}}, \bibinfo {author} {\bibfnamefont {J.~R.}\ \bibnamefont {Yates}},
  \bibinfo {author} {\bibfnamefont {G.}~\bibnamefont {Pizzi}}, \bibinfo
  {author} {\bibfnamefont {Y.-S.}\ \bibnamefont {Lee}}, \bibinfo {author}
  {\bibfnamefont {I.}~\bibnamefont {Souza}}, \bibinfo {author} {\bibfnamefont
  {D.}~\bibnamefont {Vanderbilt}}, \ and\ \bibinfo {author} {\bibfnamefont
  {N.}~\bibnamefont {Marzari}},\ }\href {\doibase
  https://doi.org/10.1016/j.cpc.2014.05.003} {\bibfield  {journal} {\bibinfo
  {journal} {Computer Physics Communications}\ }\textbf {\bibinfo {volume}
  {185}},\ \bibinfo {pages} {2309} (\bibinfo {year} {2014})}\BibitemShut
  {NoStop}%
\bibitem [{\citenamefont {Zhang}\ \emph {et~al.}(2019)\citenamefont {Zhang},
  \citenamefont {Wu}, \citenamefont {Liu},\ and\ \citenamefont
  {Yazyev}}]{MRZhangprb}%
  \BibitemOpen
  \bibfield  {author} {\bibinfo {author} {\bibfnamefont {S.}~\bibnamefont
  {Zhang}}, \bibinfo {author} {\bibfnamefont {Q.}~\bibnamefont {Wu}}, \bibinfo
  {author} {\bibfnamefont {Y.}~\bibnamefont {Liu}}, \ and\ \bibinfo {author}
  {\bibfnamefont {O.~V.}\ \bibnamefont {Yazyev}},\ }\href {\doibase
  10.1103/PhysRevB.99.035142} {\bibfield  {journal} {\bibinfo  {journal} {Phys.
  Rev. B}\ }\textbf {\bibinfo {volume} {99}},\ \bibinfo {pages} {035142}
  (\bibinfo {year} {2019})}\BibitemShut {NoStop}%
\bibitem [{\citenamefont {Liu}\ \emph {et~al.}(2009)\citenamefont {Liu},
  \citenamefont {Zhang},\ and\ \citenamefont {Yao}}]{Liuyiprb}%
  \BibitemOpen
  \bibfield  {author} {\bibinfo {author} {\bibfnamefont {Y.}~\bibnamefont
  {Liu}}, \bibinfo {author} {\bibfnamefont {H.-J.}\ \bibnamefont {Zhang}}, \
  and\ \bibinfo {author} {\bibfnamefont {Y.}~\bibnamefont {Yao}},\ }\href
  {\doibase 10.1103/PhysRevB.79.245123} {\bibfield  {journal} {\bibinfo
  {journal} {Phys. Rev. B}\ }\textbf {\bibinfo {volume} {79}},\ \bibinfo
  {pages} {245123} (\bibinfo {year} {2009})}\BibitemShut {NoStop}%
\bibitem [{\citenamefont {Wu}\ \emph {et~al.}(2018)\citenamefont {Wu},
  \citenamefont {Zhang}, \citenamefont {Song}, \citenamefont {Troyer},\ and\
  \citenamefont {Soluyanov}}]{WUWT}%
  \BibitemOpen
  \bibfield  {author} {\bibinfo {author} {\bibfnamefont {Q.}~\bibnamefont
  {Wu}}, \bibinfo {author} {\bibfnamefont {S.}~\bibnamefont {Zhang}}, \bibinfo
  {author} {\bibfnamefont {H.-F.}\ \bibnamefont {Song}}, \bibinfo {author}
  {\bibfnamefont {M.}~\bibnamefont {Troyer}}, \ and\ \bibinfo {author}
  {\bibfnamefont {A.~A.}\ \bibnamefont {Soluyanov}},\ }\href {\doibase
  https://doi.org/10.1016/j.cpc.2017.09.033} {\bibfield  {journal} {\bibinfo
  {journal} {Computer Physics Communications}\ }\textbf {\bibinfo {volume}
  {224}},\ \bibinfo {pages} {405} (\bibinfo {year} {2018})}\BibitemShut
  {NoStop}%
\bibitem [{\citenamefont {Kohler}(1938)}]{Kohlerrule}%
  \BibitemOpen
  \bibfield  {author} {\bibinfo {author} {\bibfnamefont {M.}~\bibnamefont
  {Kohler}},\ }\href {\doibase https://doi.org/10.1002/andp.19384240124}
  {\bibfield  {journal} {\bibinfo  {journal} {Annalen der Physik}\ }\textbf
  {\bibinfo {volume} {424}},\ \bibinfo {pages} {211} (\bibinfo {year}
  {1938})}\BibitemShut {NoStop}%
\bibitem [{\citenamefont {Chambers}(1952)}]{Chambers1952}%
  \BibitemOpen
  \bibfield  {author} {\bibinfo {author} {\bibfnamefont {R.~G.}\ \bibnamefont
  {Chambers}},\ }\href {\doibase 10.1088/0370-1298/65/6/114} {\bibfield
  {journal} {\bibinfo  {journal} {Proceedings of the Physical Society. Section
  A}\ }\textbf {\bibinfo {volume} {65}},\ \bibinfo {pages} {458–459}
  (\bibinfo {year} {1952})}\BibitemShut {NoStop}%
\bibitem [{sup(2024)}]{supp}%
  \BibitemOpen
  \href@noop {} {}\bibinfo {howpublished}
  {\url{URL_will_be_inserted_by_publisher}} (\bibinfo {year} {2024}),\ \bibinfo
  {note} {supplementary materials}\BibitemShut {NoStop}%
\bibitem [{\citenamefont {Zhang}\ \emph {et~al.}(2024)\citenamefont {Zhang},
  \citenamefont {Pi}, \citenamefont {Fang}, \citenamefont {Weng},\ and\
  \citenamefont {Wu}}]{HallZhang}%
  \BibitemOpen
  \bibfield  {author} {\bibinfo {author} {\bibfnamefont {S.}~\bibnamefont
  {Zhang}}, \bibinfo {author} {\bibfnamefont {H.}~\bibnamefont {Pi}}, \bibinfo
  {author} {\bibfnamefont {Z.}~\bibnamefont {Fang}}, \bibinfo {author}
  {\bibfnamefont {H.}~\bibnamefont {Weng}}, \ and\ \bibinfo {author}
  {\bibfnamefont {Q.}~\bibnamefont {Wu}},\ }\href@noop {} {\enquote {\bibinfo
  {title} {New perspectives of hall effects from first-principles
  calculations},}\ } (\bibinfo {year} {2024}),\ \Eprint
  {http://arxiv.org/abs/2401.15150} {arXiv:2401.15150 [cond-mat.mes-hall]}
  \BibitemShut {NoStop}%
\bibitem [{\citenamefont {Pi}\ \emph {et~al.}(2024)\citenamefont {Pi},
  \citenamefont {Zhang}, \citenamefont {Xu}, \citenamefont {Fang},
  \citenamefont {Weng},\ and\ \citenamefont {Wu}}]{pi2024_ZrTe5}%
  \BibitemOpen
  \bibfield  {author} {\bibinfo {author} {\bibfnamefont {H.}~\bibnamefont
  {Pi}}, \bibinfo {author} {\bibfnamefont {S.}~\bibnamefont {Zhang}}, \bibinfo
  {author} {\bibfnamefont {Y.}~\bibnamefont {Xu}}, \bibinfo {author}
  {\bibfnamefont {Z.}~\bibnamefont {Fang}}, \bibinfo {author} {\bibfnamefont
  {H.}~\bibnamefont {Weng}}, \ and\ \bibinfo {author} {\bibfnamefont
  {Q.}~\bibnamefont {Wu}},\ }\href {https://arxiv.org/abs/2401.15151} {\enquote
  {\bibinfo {title} {First-principles methodology for studying magnetotransport
  in narrow-gap semiconductors: an application to zirconium pentatelluride
  zrte5},}\ } (\bibinfo {year} {2024}),\ \Eprint
  {http://arxiv.org/abs/2401.15151} {arXiv:2401.15151 [cond-mat.mtrl-sci]}
  \BibitemShut {NoStop}%
\bibitem [{\citenamefont {Liu}\ \emph {et~al.}(2024)\citenamefont {Liu},
  \citenamefont {Zhang}, \citenamefont {Fang}, \citenamefont {Weng},\ and\
  \citenamefont {Wu}}]{zhliu2024}%
  \BibitemOpen
  \bibfield  {author} {\bibinfo {author} {\bibfnamefont {Z.}~\bibnamefont
  {Liu}}, \bibinfo {author} {\bibfnamefont {S.}~\bibnamefont {Zhang}}, \bibinfo
  {author} {\bibfnamefont {Z.}~\bibnamefont {Fang}}, \bibinfo {author}
  {\bibfnamefont {H.}~\bibnamefont {Weng}}, \ and\ \bibinfo {author}
  {\bibfnamefont {Q.}~\bibnamefont {Wu}},\ }\href
  {https://arxiv.org/abs/2401.15146} {\enquote {\bibinfo {title}
  {First-principles methodology for studying magnetotransport in magnetic
  materials},}\ } (\bibinfo {year} {2024}),\ \Eprint
  {http://arxiv.org/abs/2401.15146} {arXiv:2401.15146 [cond-mat.mtrl-sci]}
  \BibitemShut {NoStop}%
\bibitem [{\citenamefont {Sudesh}\ \emph {et~al.}(2017)\citenamefont {Sudesh},
  \citenamefont {Kumar}, \citenamefont {Neha}, \citenamefont {Das},\ and\
  \citenamefont {Patnaik}}]{Sudesh2017}%
  \BibitemOpen
  \bibfield  {author} {\bibinfo {author} {\bibnamefont {Sudesh}}, \bibinfo
  {author} {\bibfnamefont {P.}~\bibnamefont {Kumar}}, \bibinfo {author}
  {\bibfnamefont {P.}~\bibnamefont {Neha}}, \bibinfo {author} {\bibfnamefont
  {T.}~\bibnamefont {Das}}, \ and\ \bibinfo {author} {\bibfnamefont
  {S.}~\bibnamefont {Patnaik}},\ }\href {\doibase 10.1038/srep46062} {\bibfield
   {journal} {\bibinfo  {journal} {Scientific Reports}\ }\textbf {\bibinfo
  {volume} {7}} (\bibinfo {year} {2017}),\ 10.1038/srep46062}\BibitemShut
  {NoStop}%
\bibitem [{\citenamefont {Weng}\ \emph {et~al.}(2015)\citenamefont {Weng},
  \citenamefont {Fang}, \citenamefont {Fang}, \citenamefont {Bernevig},\ and\
  \citenamefont {Dai}}]{WSMNbP1}%
  \BibitemOpen
  \bibfield  {author} {\bibinfo {author} {\bibfnamefont {H.}~\bibnamefont
  {Weng}}, \bibinfo {author} {\bibfnamefont {C.}~\bibnamefont {Fang}}, \bibinfo
  {author} {\bibfnamefont {Z.}~\bibnamefont {Fang}}, \bibinfo {author}
  {\bibfnamefont {B.~A.}\ \bibnamefont {Bernevig}}, \ and\ \bibinfo {author}
  {\bibfnamefont {X.}~\bibnamefont {Dai}},\ }\href {\doibase
  10.1103/PhysRevX.5.011029} {\bibfield  {journal} {\bibinfo  {journal} {Phys.
  Rev. X}\ }\textbf {\bibinfo {volume} {5}},\ \bibinfo {pages} {011029}
  (\bibinfo {year} {2015})}\BibitemShut {NoStop}%
\bibitem [{\citenamefont {Shekhar}\ \emph {et~al.}(2015)\citenamefont
  {Shekhar}, \citenamefont {Nayak}, \citenamefont {Sun}, \citenamefont
  {Schmidt}, \citenamefont {Nicklas}, \citenamefont {Leermakers}, \citenamefont
  {Zeitler}, \citenamefont {Skourski}, \citenamefont {Wosnitza}, \citenamefont
  {Liu}, \citenamefont {Chen}, \citenamefont {Schnelle}, \citenamefont
  {Borrmann}, \citenamefont {Grin}, \citenamefont {Felser},\ and\ \citenamefont
  {Yan}}]{WSMNbP2}%
  \BibitemOpen
  \bibfield  {author} {\bibinfo {author} {\bibfnamefont {C.}~\bibnamefont
  {Shekhar}}, \bibinfo {author} {\bibfnamefont {A.~K.}\ \bibnamefont {Nayak}},
  \bibinfo {author} {\bibfnamefont {Y.}~\bibnamefont {Sun}}, \bibinfo {author}
  {\bibfnamefont {M.}~\bibnamefont {Schmidt}}, \bibinfo {author} {\bibfnamefont
  {M.}~\bibnamefont {Nicklas}}, \bibinfo {author} {\bibfnamefont
  {I.}~\bibnamefont {Leermakers}}, \bibinfo {author} {\bibfnamefont
  {U.}~\bibnamefont {Zeitler}}, \bibinfo {author} {\bibfnamefont
  {Y.}~\bibnamefont {Skourski}}, \bibinfo {author} {\bibfnamefont
  {J.}~\bibnamefont {Wosnitza}}, \bibinfo {author} {\bibfnamefont
  {Z.}~\bibnamefont {Liu}}, \bibinfo {author} {\bibfnamefont {Y.}~\bibnamefont
  {Chen}}, \bibinfo {author} {\bibfnamefont {W.}~\bibnamefont {Schnelle}},
  \bibinfo {author} {\bibfnamefont {H.}~\bibnamefont {Borrmann}}, \bibinfo
  {author} {\bibfnamefont {Y.}~\bibnamefont {Grin}}, \bibinfo {author}
  {\bibfnamefont {C.}~\bibnamefont {Felser}}, \ and\ \bibinfo {author}
  {\bibfnamefont {B.}~\bibnamefont {Yan}},\ }\href {\doibase 10.1038/nphys3372}
  {\bibfield  {journal} {\bibinfo  {journal} {Nature Physics}\ }\textbf
  {\bibinfo {volume} {11}},\ \bibinfo {pages} {645} (\bibinfo {year}
  {2015})}\BibitemShut {NoStop}%
\bibitem [{\citenamefont {Leahy}\ \emph {et~al.}(2018)\citenamefont {Leahy},
  \citenamefont {Lin}, \citenamefont {Siegfried}, \citenamefont {Treglia},
  \citenamefont {Song}, \citenamefont {Nandkishore},\ and\ \citenamefont
  {Lee}}]{kohlerNbP}%
  \BibitemOpen
  \bibfield  {author} {\bibinfo {author} {\bibfnamefont {I.~A.}\ \bibnamefont
  {Leahy}}, \bibinfo {author} {\bibfnamefont {Y.-P.}\ \bibnamefont {Lin}},
  \bibinfo {author} {\bibfnamefont {P.~E.}\ \bibnamefont {Siegfried}}, \bibinfo
  {author} {\bibfnamefont {A.~C.}\ \bibnamefont {Treglia}}, \bibinfo {author}
  {\bibfnamefont {J.~C.~W.}\ \bibnamefont {Song}}, \bibinfo {author}
  {\bibfnamefont {R.~M.}\ \bibnamefont {Nandkishore}}, \ and\ \bibinfo {author}
  {\bibfnamefont {M.}~\bibnamefont {Lee}},\ }\href {\doibase
  10.1073/pnas.1808747115} {\bibfield  {journal} {\bibinfo  {journal}
  {Proceedings of the National Academy of Sciences}\ }\textbf {\bibinfo
  {volume} {115}},\ \bibinfo {pages} {10570} (\bibinfo {year}
  {2018})}\BibitemShut {NoStop}%
\end{thebibliography}%
\bibliographystyle{apsrev4-1}      

\end{document}